# Magnetically Modulated Electrical Switching in an Antiferromagnetic Transistor

Chung-Tao Chou[1,2], Eugene Park[3], Josep Ingla-Aynés[4], Julian Klein[3], Kseniia Mosina[5], Jagadeesh S. Moodera[1,4], Zdenek Sofer[5], Frances M. Ross[3], Luqiao Liu[2*]

[1]Department of Physics, Massachusetts Institute of Technology, Cambridge, MA 02139, USA.

[2]Department of Electrical Engineering and Computer Science, Massachusetts Institute of Technology, Cambridge, MA 02139, USA.

[3]Department of Materials Science and Engineering, Massachusetts Institute of Technology, Cambridge, MA 02139, USA.

[4]Francis Bitter Magnet Laboratory and Plasma Science and Fusion Center, Massachusetts Institute of Technology, Cambridge, Massachusetts 02139, USA.

[5]Department of Inorganic Chemistry, University of Chemistry and Technology Prague, Technická 5, Prague 166 28, Czech Republic.

**Abstract**

A spin version of transistor, where magnetism is used to influence electrical behaviors of the semiconductor, has been a long-pursued device concept in spintronics. In this work, we experimentally study a field-effect transistor with CrSBr, a van der Waals (vdW) antiferromagnetic semiconductor, as the channel material. Unlike the weak magnetic tunability of in-plane currents previously reported in vdW magnets, the channel current of our transistor is efficiently tuned by both gate voltage and magnetic transitions, achieving a magnetoresistance ratio as high as 1500%. Combining measurement and theoretical modeling, we reveal magnetically modulated carrier concentration as the origin of the large magnetoresistance. The strategy of using both magnetic ordering and electric field in the same device to control ON/OFF states of a transistor opens a new avenue of energy-efficient spintronics for memory, logic and magnetic sensing applications.

The metal-oxide-semiconductor field effect transistor (MOSFET) is one of the most important building blocks in current electronic systems. An electric field applied across an insulating dielectric controls the conductance of a semiconductor through the Coulomb interaction with the electron charge. Spintronic devices, on the other hand, exploit the spin degree of freedom of electrons to enable memory, logic, sensing and other functions. Since the birth of spintronics, a goal has been to incorporate spin or magnetism into semiconductors and realize a spin version of the transistor [1,2], where the spin orientation switches the electrical current ON and OFF. Such a device can act not only as a basic component for spin logic, but also potentially help to address leakage current or subthreshold swing issue in standard MOSFET by providing sharper and more energy efficient switching through the collective dynamics of magnetic ordering [3,4]. One well-known example of these proposed devices is the Datta-Das spin transistor [5]. However, because of the stringent requirements on the efficiency of spin injection and detection at the semiconductor/magnet interface [6] and the need for coherent spin transport [7], the realization of such devices with high ON/OFF ratio in a local geometry remains elusive.

Two-dimensional (2D) van der Waals (vdW) magnetic materials show rich optical, magnetic, and electrical properties [8-12]. Many of these materials exhibit semiconducting behavior due to a moderate band gap, and simultaneously show relatively weak interlayer antiferromagnetic interaction that leads to various magnetic transitions including spin-flip, spin-flop, and spin-canting [13-15] when subjected to a magnetic field, exchange field [16], or spin torque [17,18]. It has recently been suggested from first-principles calculations and optical measurements that, aside from its direct influence on magnetic properties, the spin transition in these materials can also affect the electronic band structure due to an exchange interaction between local magnetic atoms and conduction electrons [19-22]. As the magnetic orientation of adjacent atomic layers changes from antiparallel (AP) to parallel (P), the originally forbidden hopping of electrons with opposite spins between neighboring layers becomes allowed, leading to a sizable reduction in the bandgap [Fig. 1(e)]. This band structure change therefore provides an additional means, on top of the electrical field effect, to magnetically tune the carrier concentration as well as the resultant current in a MOSFET.

Experimental studies have recently been carried out to probe the effect of magnetic field on transport within antiferromagnetic semiconductors [23]. However, only very weak magnetic modulation on the in-plane current has been observed [23, 24], several orders of magnitude lower than the tunneling or hopping currents in vertical tunnel junction structures [17, 25]. In addition, these in-plane devices also often exhibit weak gate voltage tunability [23]. The low modulation efficiency with magnetic and electrical controls greatly compromises their function as useful magnetic transistor devices, and hampers understandings of underlying physics [23,24]. In this work, we realize a magnetic MOSFET [Fig. 1(a)] where the channel current is efficiently modulated not only by a gate voltage (ON/OFF ratio > $10^6$) but also by the transition between antiferromagnetic and ferromagnetic states, with a magnetoresistance of ~1500% at 35 K under sharp magnetic switching. Through experiment and theoretical modeling, we further reveal the spin-flip-induced carrier concentration variation as the key source of magnetic control over the channel conductance [Fig. 1(f)-(g)], a mechanism that greatly enhances magnetic tunability over existing magnetoresistance effects. Moreover, by correlating the transport properties with magnetic transition behavior, we demonstrate that the magnetic MOSFET is a suitable platform for identifying and distinguishing interfacial magnetic switching property from bulk behavior.

We fabricate the transistor [Fig. 1(a)] using single crystalline CrSBr as the channel layer. CrSBr is known for its moderate bandgap of ~1.5 eV [26] and its sharp spin-flip transition observable up to 130 K [13]. CrSBr grown by chemical vapor transport is mechanically exfoliated, and flakes with a thickness of typically a few hundred nm are transferred onto pre-patterned source and drain electrodes on oxidized silicon substrate (see Supplemental Material and Fig. S1(a)-(c) [27]). The silicon substrate is heavily doped and used as a back gate electrode. Since the oxide/semiconductor interface is critical for MOSFET performance, we aim to minimize damage and contamination at this interface by optimizing passivation and transfer steps [27]. The quality of the pristine CrSBr was assessed by X-ray diffraction (XRD) [Fig. S1(d)], while the interface structure within the devices was examined using cross-sectional scanning transmission electron microscopy (STEM) [Fig. 1(a)-(b)]. The magnetism of CrSBr is characterized by vibrating sample magnetometry and a clear spin-flip transition is observed below the Néel temperature $T_N$ = 130 K [Fig. 1(c)-(d)] [13].

We first measure the temperature dependence of the drain current ($I_{DS}$) of the magnetic transistor under fixed drain-to-source ($V_{DS}$) and gate-to-source voltages ($V_{GS}$). A sharp dip in the second derivative of electrical conductance with respect to temperature is observed at $T$ = 130 K [Fig. S2(a)], an indication of the clean Néel transition from the paramagnetic to the antiferromagnetic state [23], consistent with the magnetization characterization. We further characterize the basic transistor properties of the device by measuring $I_{DS}$ under various $V_{GS}$ [Fig. 2(a)] and $V_{DS}$ [Fig. 2(b)]. The $I_{DS} - V_{GS}$ results show typical n-MOSFET curves with ON/OFF ratio of more than $10^6$. The threshold voltage ($V_{th}$) decreases with increasing temperature [Fig. S2(b)], which is typical in MOSFETs with a lightly doped body where the dopants freeze out at low temperatures [28].

The high ON/OFF ratio under $V_{GS}$ is one key difference exhibited between our devices and previous studies on CrSBr [23,24,29,30]. In the latter, CrSBr often suffers from surface defects and background dopants and therefore the tunable range of carrier density is very limited. The high electrical tunability of our sample enables characterization of magnetic modulation on transport over a much wider range of carrier densities and enables us to observe the drastically different magnetoresistance in different carrier density regimes, as will be shown later. The high gate voltage tunability could be related to the reduced density of states of interface trap $D_{it}$, which generally screens out the electrical field effect. Extracting from the subthreshold slope in Fig. 2(a), we obtain $D_{it}$ ~$10^{13}$ cm$^{-2}$eV$^{-1}$ [27], comparable to typical transistors made with transition metal dichalcogenides [31], and at least an order of magnitude lower than the corresponding values in previous CrSBr devices.

We study the magnetic modulation of the electrical transport behavior of the magnetic transistor by applying a magnetic field along the easy axis (b-axis) of CrSBr and measuring magnetoresistance at different temperatures. We observe clear switching at the spin-flip transition fields below the Néel temperature [Fig. 2(c)]. The influence of the gate voltage on the magnetoresistance is further reflected in the $I_{DS} - V_{GS}$ characteristics between AP and P states [Fig. 2(d) and Fig. S2(c)], where the applied magnetic field induces a noticeable splitting in the $I_{DS}$ curve with a shift in threshold voltage. Following the conventions of magnetoresistance, we define the channel magnetoresistance (MR) as $(I_{DS}^{P} - I_{DS}^{AP})/I_{DS}^{AP}$ and summarize the results in Fig. 3(a). We find that the MR exhibits a strong gate voltage dependence: it is largest when the device is in the deep subthreshold regime but decreases and saturates at ~30% when the electron channel is formed under a positive $V_{GS}$. Qualitatively consistent results are observed in multiple devices [Fig. 3(c) and Fig. S3]. Under high $V_{GS}$, the obtained relatively low MR (30%) is similar to previous studies on CrSBr [23,24,29,30], which all share a high conductivity and large carrier concentration. By accessing the ultralow carrier density region in deep subthreshold, our device reveals the strong carrier-density-dependence and achieves a much higher ratio of MR [Fig. 3(a)].

It is well known that vdW magnets like $CrI_3$ and CrSBr can exhibit large tunneling or hopping magnetoresistance in vertical junction structures [25, 32-35]. By performing finite element transport simulations accounting for anisotropic charge transport, we show that the magnetoresistance observed in our experiment *cannot* be attributed to this mechanism (Fig. S4). Specifically, the large increase of MR with the decrease of $V_{GS}$ in Fig. 3(a) shows that the magnetoresistance becomes more significant as the channel resistance increases, opposite to the trend one would expect from a picture where the contact tunnel junctions dominate. Additionally, unlike tunneling magnetoresistance (TMR) which typically shows a strong dependence on bias voltage [25, 32], the large MR in our measurement exhibits only a weak dependence on the applied $V_{DS}$ [Fig. S2(d)]. Finally, our quantitative simulation shows that the channel current responds weakly to the toggling of vertical junction resistance [Fig. S4(c)], highlighting the insignificant contribution from TMR.

The different behaviors of MR across the two regimes of transistor operation requires further analysis of the magnetic influence on the semiconductor electrostatics. In the electrical ON regime (the triode inversion regime in nMOS), the drain current, which flows in b-axis direction, follows the relationship $I_{DS}^{P,AP} = \mu_{n,b}^{P,AP} \frac{W}{L} C_{ox} \left(V_{GS} - V_{th} - \frac{1}{2}V_{DS}\right) V_{DS}$ [36], where $W$, $L$, $C_{ox}$, and $\mu_{n,b}^{P,AP}$ represent channel width, length, gate oxide capacitance and electron mobility along the b-axis in the P or AP state, separately. In this case, it is known that the carrier density inside the conducting channel is dominated by the gate capacitance $C_{ox}$. Therefore, the distinct slopes of $I_{DS}$ for AP and P states of Fig. 2(d) mainly reflect the mobility difference between the two magnetic states. By fitting the straight segments of Fig. 2(d), we extract mobility values of $\mu_{n,b}^{AP}$ = 8 cm²V⁻

$^{-1}$s$^{-1}$ and $\mu_{n,b}^{P}$ = 11 cm$^2$V$^{-1}$s$^{-1}$. The ~30% mobility difference between the AP and P states is potentially related to the spin-dependent scattering at layer interfaces, similar to the well-known current-in-plane (CIP) giant magnetoresistance (GMR) effect [37,38]. We note that these mobility values are significantly higher than those in monolayer and bilayer CrSBr [23], likely due to the larger flake thickness and lower interface trap density.

The large MR in the subthreshold regime is from the combined effects of the magnetic modulations on the mobility and on carrier densities. For a standard MOSFET, the drain current in the subthreshold regime is given by $I_{DS}^{P,AP} \approx \frac{W}{L} k_B T \mu_{n,b}^{P,AP} n^{P,AP}$ [36], where $k_B$ and $T$ are the Boltzmann constant and temperature and $n^{P,AP}$ represents the carrier concentration in the two states. Modeling a MOSFET with thin semiconductor body, we derive the ratio of the drain subthreshold currents in the two magnetic states (see Supplementary Text and Fig. S5 [27]):

$$\frac{I_{DS}^{P}}{I_{DS}^{AP}} \approx \frac{1}{2} \frac{\mu_{n,b}^{P}}{\mu_{n,b}^{AP}} e^{\frac{\Delta E_C}{k_B T}} \qquad (1),$$

where $\Delta E_C = E_C^{AP} - E_C^{P}$ is the energy shift of conduction band edge with respect to the Fermi level. Since the measurement in the electrical ON regime shows $\mu_{n,b}^{P,AP}$ differ only by about 30%, the large MR ratio in the subthreshold should therefore be mostly accounted for by the carrier concentration variation through $\Delta E_C$. As a consequence of the Boltzmann distribution, Eq. (1) suggests that the most prominent feature of this mechanism is the exponential dependence of the MR on the ratio between $\Delta E_C$ and thermal energy $k_B T$. To verify this, we measure under different temperatures [Fig. 3(a)] and find that the MR curves at different temperatures in Fig. 3(a) can be universally reproduced from our model with the same parameter of $\Delta E_C$ = 9 meV [Fig. 3(b)-(c)]. The measured $\Delta E_C$ value roughly aligns with the bandgap change ($\Delta E_g = \Delta E_C + \Delta E_V$, $\Delta E_V$ being the valence band edge shift) estimated from the exciton spectroscopy data (~20 meV) [19], but lies below the theoretically predicted value (~0.1 eV) [39,40]. Further study is required to determine the exact relation between $\Delta E_g$ and $\Delta E_C$, which can depend on factors like the type and density of background dopants [36].

As the spin transition can control the ON/OFF states under a fixed voltage, we introduce a new figure of merit, magnetic subthreshold swing $SS_{mag}$, analogous to the electrical subthreshold swing, to quantify the energy efficiency:

$$SS_{mag} = \mu_B \frac{\Delta H}{\Delta (\log_{10} I_{DS})} \qquad (2),$$

where $\mu_B$ is the Bohr magneton and $\Delta H$ is the magnetic field required to switch between magnetic states. The $SS_{mag}$ value in our device is ~2 μeV/decade at 40 K [Fig. 2(c)], more than three orders of magnitude smaller than the theoretical limit of electrical subthreshold swing (8 meV/decade for this temperature). We note that in previous studies of other antiferromagnets such as CrPS$_4$ [41], $SS_{mag}$, when calculated using Eq. 2, is comparable to the electrical subthreshold swing limit under corresponding temperatures, due to the high required magnetic field (~10 Tesla). Besides cutting into the energy efficiency, the high field requirement in CrPS$_4$ also poses difficulties in distinguishing from other effects such as chiral anomaly or weak localization, which exhibit similar magnitudes over a comparable field range [42,43]. The sub-thermal magnetic subthreshold swing is enabled by the collective switching of magnetic ordering rather than single electron behavior, similar to the steep switching mechanisms in the phase-transition hyper FET [44] and ferroelectric FETs [45]. Materials with sharper switching and larger bandgap difference between P and AP states could further improve the energy efficiency.

During our MR measurement, we find that $I_{DS}$ vs. $H$ curves exhibit different detailed characteristics under different $V_{GS}$. As in Fig. 4(a), deep in the subthreshold regime ($V_{GS}$ = -5 V), the MR change occurs sharply at a single transition field $H_{sp}$, resembling the M-H curve [Fig. 1(d)]. Meanwhile, when the device is close to electrical ON regime ($V_{GS}$ = 2 V and 5 V in Fig. 4(a)), the magnetoresistance exhibits a two-step transition, with the two transition fields $H_{sp1} \approx H_{sp}$ and $H_{sp2} \approx \frac{1}{2} H_{sp}$, between which a resistance plateau develops (yellow shaded region in Fig. 4(a)) and becomes more pronounced as $V_{GS}$ increases. The distinct $I_{DS}$ vs. H characteristics in electrical ON regime is summarized in Fig. 4(b), where the intermediate state region ('inter*') is labelled.

The emergence of two-step transition only appears in the electrical transport measurement and is absent in the corresponding M-H curve. It is well known that a nanometer-thin channel layer forms when a MOSFET enters the electrical ON regime, where the carriers become heavily concentrated at the oxide/semiconductor interface [Fig. S6]. Therefore, $I_{DS}$ in electrical ON region represents more of the property local to the interface than that of the body. We calculate the energies of different magnetic arrangements [Fig. 4(c)] and show that the magnetic layer of CrSBr nearest the surface is expected to switch under a lower magnetic field ($\sim \frac{1}{2} H_{sp}$) due to the existence of only one nearest neighbor, in contrast to the two neighbors in the bulk [19,46]. Using the calculated carrier distribution, we estimate the height of the plateau in $I_{DS}$, which agrees well with the experiment [Fig. 4(d)]. This unique sensitivity to the magnetic switching in interfacial layers allows us to distinguish the different magnetic properties between surface and bulk.

In conclusion, we describe and operate magnetic MOSFET structures where the magnetism is used to control the electrical switching. Unlike mechanisms such as anisotropic magnetoresistance, GMR and TMR, which modify the scattering time [37,38] or tunneling matrix [32,33] of carriers, the magnetic state directly acts on the concentration of carriers in our revealed effect, possessing the potential of achieving a giant magnitude. In the current measurement, we demonstrate spin transition-induced magnetoresistance at cryogenic temperature. However, with the revealed physics, we anticipate that the same effect could in principle also work at higher or even room temperature, if antiferromagnetic materials with high Néel temperature [47] were to be employed. The high magnetic sensitivity of the channel resistance can be directly utilized for magnetic field sensing, where the large magnetoresistance is beneficial for improving the signal-to-noise ratio, and simultaneously the gate voltage can be used to fine-tune the working condition of the device. Aside from controlling with magnetic field, electrical means such as spin torque [17] or voltage induced exchange interaction change [48] can be employed to induce spin transitions, useful for locally switching the transistors. Combining the electrical and magnetic degrees of freedom, the magnetic MOSFET will make compact designs for spin logic or magnetic memory applications.

*Acknowledgements*—This work was supported in part by Semiconductor Research Corporation (SRC) and DARPA, National Science Foundation under award DMR-2104912. E.P., J.K., and F.R. were supported by Department of Energy grant DE-SC0025387. The magnetometry measurements are supported by the National Science Foundation (NSF-DMR 2218550) and the Army Research Office (W911NF-20-2-0061, DURIP W911NF-20-1-0074). Z.S. and K.M. were supported by ERC-CZ program (project LL2101) from Ministry of Education Youth and Sports (MEYS) and by the project Advanced Functional Nanorobots (reg. No. CZ.02.1.01/0.0/0.0/15_003/0000444 financed by the ERDF).

C.T.C. and E.P. are designated as co-first authors. C.T.C. performed transport measurements, theoretical modeling, and manuscript writing. E.P. performed TEM imaging.

*Corresponding author. Email: [luqiao@mit.edu](luqiao@mit.edu)

# References

[1] Sugahara, S. & Nitta, J. Spin-transistor electronics: An overview and outlook. *Proceedings of the IEEE* **98**, 2124–2154 (2010).

[2] Dietl, T. & Ohno, H. Dilute ferromagnetic semiconductors: Physics and spintronic structures. *Reviews of Modern Physics* **86**, 187–251 (2014).

[3] Manipatruni, S., Nikonov, D. E. & Young, I. A. Beyond CMOS computing with spin and polarization. *Nature Physics* **14**, 338–343 (2018).

[4] Manipatruni, S. *et al.* Scalable energy-efficient magnetoelectric spin–orbit logic. *Nature* **565**, 35–42 (2019).

[5] Datta, S. & Das, B. Electronic analog of the electro-optic modulator. *Applied Physics Letters* **56**, 665–667 (1990).

[6] Lou, X. *et al.* Electrical detection of spin transport in lateral ferromagnet–semiconductor devices. *Nature Physics* **3**, 197–202 (2007).

[7] Koo, H. C. *et al.* Control of spin precession in a spin-injected field effect transistor. *Science* **325**, 1515–1518 (2009).

[8] Huang, B. *et al.* Layer-dependent ferromagnetism in a van der Waals crystal down to the monolayer limit. *Nature* **546**, 270–273 (2017).

[9] Gong, C. *et al.* Discovery of intrinsic ferromagnetism in two-dimensional van der Waals crystals. *Nature* **546**, 265–269 (2017).

[10] Gao, F. Y. *et al.* Giant chiral magnetoelectric oscillations in a van der Waals multiferroic. *Nature* **632**, 273-279 (2024).

[11] Sun, Z. *et al.* Giant nonreciprocal second-harmonic generation from antiferromagnetic bilayer $CrI_3$. *Nature* **572**, 497–501 (2019).

[12] Chen, Y. *et al.* Twist-assisted all-antiferromagnetic tunnel junction in the atomic limit. *Nature* **632**, 1045-1051 (2024).

[13] López-Paz, S. A. *et al.* Dynamic magnetic crossover at the origin of the hidden-order in van der Waals antiferromagnet CrSBr. *Nature Communications* **13**, 4745 (2022).

[14] Huang, B. *et al.* Electrical control of 2D magnetism in bilayer $CrI_3$. *Nature Nanotechnology* **13**, 544–548 (2018).

[15] Morrison, B. The spin-flop transition in some two-sublattice uniaxial antiferromagnets.

*physica status solidi (b)* **59**, 581–587 (1973).

[16] Stamps, R. Mechanisms for exchange bias. *Journal of Physics D: Applied Physics* **33**, R247 (2000).

[17] Cham, T. M. J. *et al.* Spin-filter tunneling detection of antiferromagnetic resonance with electrically-tunable damping. *arXiv preprint arXiv:2407.09462* (2024).

[18] Manchon, A. *et al.* Current-induced spin-orbit torques in ferromagnetic and antiferromagnetic systems. *Reviews of Modern Physics* **91**, 035004 (2019).

[19] Wilson, N. P. *et al.* Interlayer electronic coupling on demand in a 2D magnetic semiconductor. *Nature Materials* **20**, 1657–1662 (2021).

[20] Bae, Y. J. *et al.* Exciton-coupled coherent magnons in a 2D semiconductor. *Nature* **609**, 282–286 (2022).

[21] Sun, Y. *et al.* Dipolar spin wave packet transport in a van der Waals antiferromagnet. *Nature Physics* **20**, 794-800 (2024).

[22] Santos-Cottin, D. *et al.* $EuCd_2As_2$: A magnetic semiconductor. *Physical Review Letters* **131**, 186704 (2023).

[23] Telford, E. J. *et al.* Coupling between magnetic order and charge transport in a two-dimensional magnetic semiconductor. *Nature Materials* **21**, 754–760 (2022).

[24] Telford, E. J. *et al.* Layered antiferromagnetism induces large negative magnetoresistance in the van der Waals semiconductor CrSBr. *Advanced Materials* **32**, 2003240 (2020).

[25] Lin, X. *et al.* Influence of magnetism on vertical hopping transport in CrSBr. *Physical Review Research* **6**, 013185 (2024).

[26] Ziebel, M. E. *et al.* CrSBr: an air-stable, two-dimensional magnetic semiconductor. *Nano Letters* **24**, 4319–4329 (2024).

[27] See Supplemental Material (URL to be added) for experimental methods, theoretical modeling, and supplementary figures (Fig. S1-S6).

[28] Balestra, F. & Ghibaudo, G. Brief review of the MOS device physics for low temperature electronics. *Solid-State Electronics* **37**, 1967–1975 (1994).

[29] Jo, J. *et al.* Nonvolatile electric control of antiferromagnet CrSBr. *Nano Letters* **24**, 4471–4477 (2024).

[30] Ye, C. et al. Layer-dependent interlayer antiferromagnetic spin reorientation in air-stable semiconductor CrSBr. *ACS Nano* **16**, 11876–11883 (2022).

[31] Kim, H.-J., Kim, D.-H., Jeong, C.-Y., Lee, J.-H. & Kwon, H.-I. Determination of interface and bulk trap densities in high-mobility p-type $WSe_2$ thin-film transistors. *IEEE Electron Device Letters* **38**, 481–484 (2017).

[32] Song, T. *et al.* Giant tunneling magnetoresistance in spin-filter van der Waals heterostructures. *Science* **360**, 1214–1218 (2018).

[33] Klein, D. R. *et al.* Probing magnetism in 2D van der Waals crystalline insulators via electron tunneling. *Science* **360**, 1218–1222 (2018).

[34] Boix-Constant, C. et al. Probing the spin dimensionality in single-layer CrSBr van der Waals heterostructures by magneto-transport measurements. *Advanced Materials* **34**, 2204940 (2022).

[35] Chen, Y. et al. Twist-assisted all-antiferromagnetic tunnel junction in the atomic limit. *Nature* **632**, 1045–1051 (2024).

[36] Sze, S. M., Li, Y. & Ng, K. K. *Physics of Semiconductor Devices* (John wiley & sons, 2021).

[37] Binasch, G., Grünberg, P., Saurenbach, F. & Zinn, W. Enhanced magnetoresistance in layered magnetic structures with antiferromagnetic interlayer exchange. *Physical Review B* **39**, 4828 (1989).

[38] Baibich, M. N. *et al.* Giant magnetoresistance of (001) Fe/(001) Cr magnetic superlattices. *Physical Review Letters* **61**, 2472 (1988).

[39] Klein, J. *et al.* The bulk van der Waals layered magnet CrSBr is a quasi-1D material. *ACS Nano* **17**, 5316–5328 (2023).

[40] Heißenbüttel, M.-C. *et al.* Quadratic optical response of CrSBr controlled by spin-selective interlayer coupling. *arXiv preprint arXiv:2403.20174* (2024).

[41] Wu, F. *et al.* Gate-controlled magnetotransport and electrostatic modulation of magnetism in 2D magnetic semiconductor $CrPS_4$. *Advanced Materials* **35**, 2211653 (2023).

[42] Xiong, J. *et al.* Evidence for the chiral anomaly in the Dirac semimetal $Na_3Bi$. *Science* **350**, 413–416 (2015).

[43] Breunig, O. *et al.* Gigantic negative magnetoresistance in the bulk of a disordered topological insulator. *Nature Communications* **8**, 15545 (2017).

[44] Shukla, N. *et al.* A steep-slope transistor based on abrupt electronic phase transition. *Nature Communications* **6**, 7812 (2015).

[45] Ko, E., Lee, J. W. & Shin, C. Negative capacitance FinFET with sub-20-mV/decade subthreshold slope and minimal hysteresis of 0.48 v. *IEEE Electron Device Letters* **38**, 418–421 (2017).

[46] Ge, W. *et al.* Direct visualization of surface spin-flip transition in MnBi$_4$Te$_7$. *Physical Review Letters* **129**, 107204 (2022).

[47] Han, R., Xue, X. & Li, P. Enhanced ferromagnetism, perpendicular magnetic anisotropy and high Curie temperature in the van der Waals semiconductor CrSeBr through strain and doping. *Physical Chemistry Chemical Physics* **26**, 12219–12230 (2024).

[48] Jiang, S., Shan, J. & Mak, K. F. Electric-field switching of two-dimensional van der Waals magnets. *Nature Materials* **17**, 406–410 (2018).

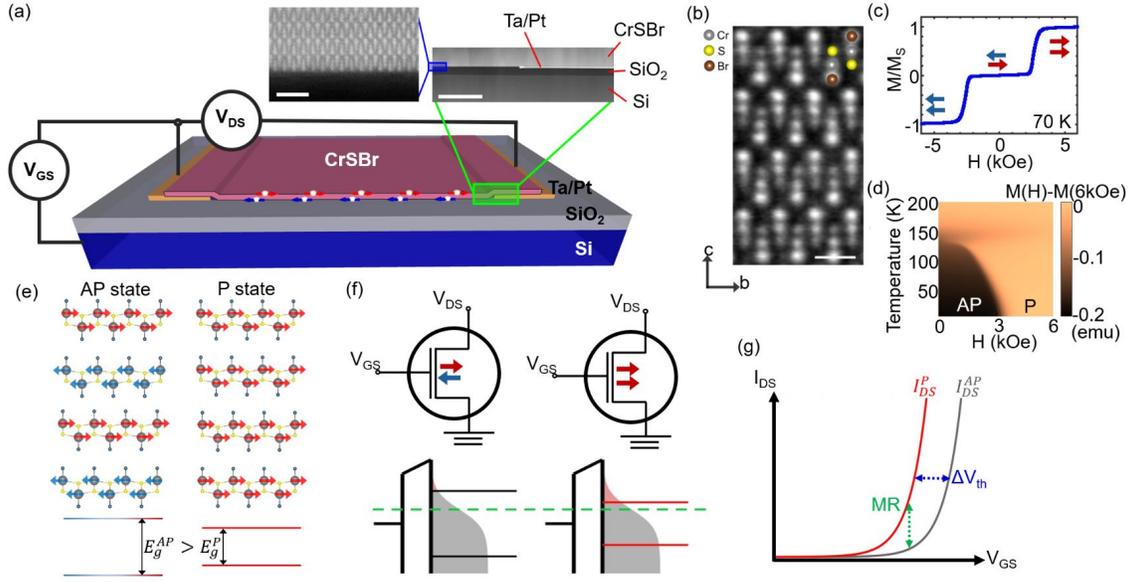

FIG. 1. (a) Schematic of the device and the measurement circuit. Insets: cross sectional STEM images viewed in the a-axis under high- (left) and low (right) magnifications. The scale bars represent 2 nm (left) and 500 nm (right). (b) High-resolution cross-sectional STEM image of CrSBr viewed in the a-axis. Scale bar 0.5 nm. (c) Magnetization curve of CrSBr measured at 70 K with field applied along the b-axis. (d) Magnetic moment change as a function of applied magnetic field at different temperatures. (e) Schematics of crystalline and magnetic structure of CrSBr before and after spin-flip transition. (f) Schematics of magnetic MOSFET in the AP and P states. The green dashed line represents the Fermi level and gray shaded regions indicate the Fermi-Dirac distribution. (g) Schematics of $I_{DS}$ as a function of $V_{GS}$ in AP and P states. The threshold voltage shift $\Delta V_{th}$ and channel magnetoresistance are expected as a result of the band gap difference shown in (e).

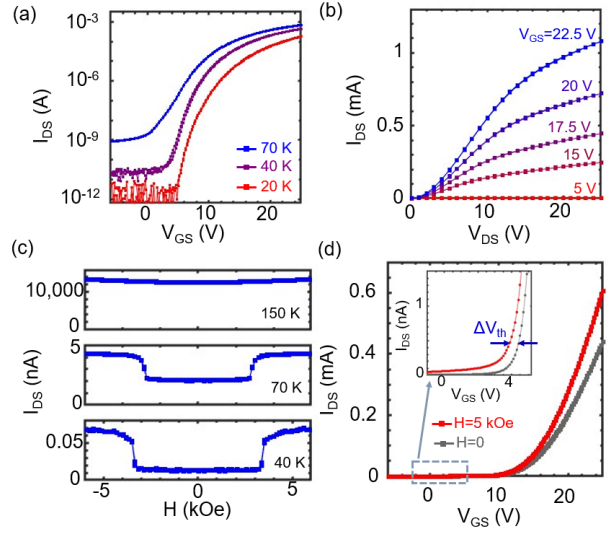

FIG. 2. (a) $I_{DS}$ vs. $V_{GS}$ at different temperatures. (b) $I_{DS}$ vs. $V_{DS}$ under different $V_{GS}$ at 70 K. (c) $I_{DS}$ vs. magnetic field $H$ at different temperatures. $V_{GS}$ is 0 V. (d) $I_{DS}$ vs. $V_{GS}$ at 40 K with $H = 0$ Oe (black) and 5 kOe (red). Inset: a magnified view in the subthreshold region. $V_{DS}$ set to be 10 V for (a), (c), and (d).

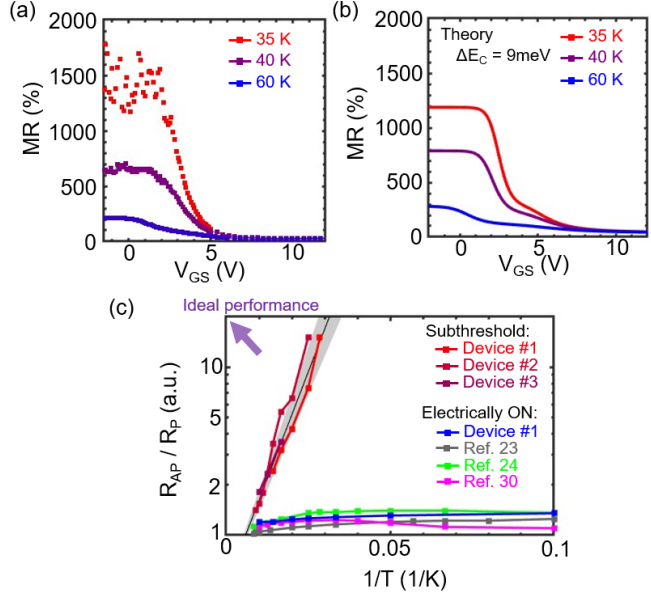

FIG. 3. (a) Magnetoresistance (MR) ratio versus $V_{GS}$ under different temperatures. $V_{DS}$ is fixed at 10 V. (b) MR calculated using the MOSFET model. A universal $\Delta E_C = 9$ meV is used. (c) Comparison of in-plane MR ratio in this work (device #1, #2, and #3) vs. previously reported results in CrSBr. The latter possess similar behaviors with our device in the transistor ON regime. The solid straight line represents the theoretical values for subthreshold regime from Eq. (1) with $\Delta E_C = 9\pm1$ meV, with the gray shaded area representing the error bar.

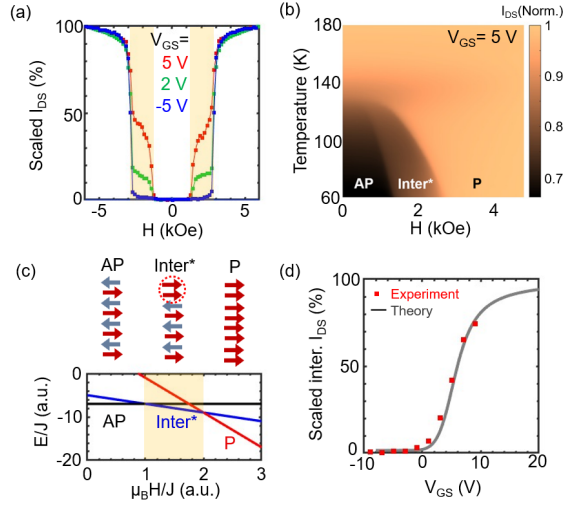

FIG. 4. (a) $I_{DS}$ vs. magnetic field $H$ under different gate voltages. Here $I_{DS}$ is normalized as $(I_{DS}(H) - I_{DS}^{AP})/(I_{DS}^{P} - I_{DS}^{AP})$. (b) $I_{DS}$ as a function of $H$ and temperature for $V_{GS} = 5$ V. (c) Magnetic configuration and relative magnetic energy in AP, intermediate, and P states versus applied magnetic field. Here, the total energy $E$ and Zeeman energy are scaled with the interlayer exchange energy $J$. (d) $I_{DS}$ in the intermediate state as a function of $V_{GS}$. Here the scaled $I_{DS}$ is defined as $(I_{DS}^{inter} - I_{DS}^{AP})/(I_{DS}^{P} - I_{DS}^{AP})$. The experimental results (red squares) are compared with theory (gray line), as calculated using the carrier concentration in the two layers nearest to the interface.

**End matters**

We also systematically carry out magnetoresistance measurements with different field direction ($H \parallel$ a-axis or b-axis, see Fig. 5(a)) and current flowing direction ($I_{DS} \parallel$ a-axis or b-axis) and find consistent results with our theoretical model. First, comparing results with field applied along different directions, the MR only depends on the relative canting angle of the adjacent layer as expected [Fig. 5(b)]. In addition, we found that MR has different dependences on current flowing direction in the two operation regimes. In the subthreshold regime, the MR is consistently large, independent of $I_{DS}$ flowing direction [Fig. 5(c)], agreeing with the carrier density modulation picture as the carrier density is equally influenced by the magnetic state in these two cases. On the other hand, the MR in the electrical ON state exhibits different values with $I_{DS}$ parallel to a-axis (2%) or b-axis (30%) [Fig. 5(d)], suggesting the different influences of magnetic state onto the mobility values for these two orientations [Fig. 5(e) and 5(f)]. We note that the magnetically independent parts of the mobility along the two axes are very different to start with [Fig. 5(f)], due to the anisotropic crystal structures.

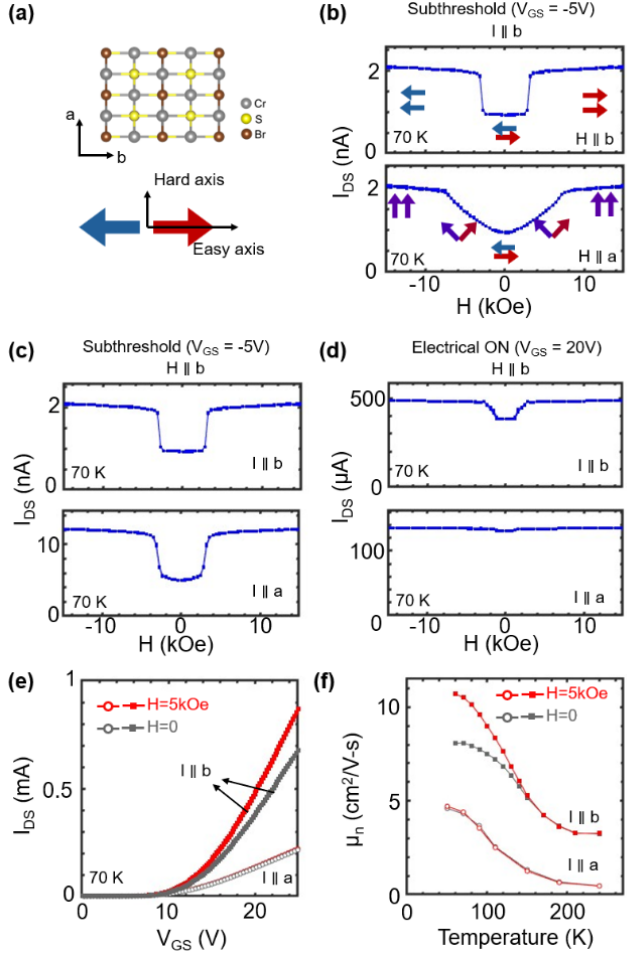

FIG. 5. (a) Anisotropic crystal structure of CrSBr and its magnetic easy and hard axes. (b) $I_{DS}$ as a function of the applied field along the magnetic easy (top) and hard (bottom) axes, $H \parallel b$ and $H \parallel a$, respectively. The arrows indicate the magnetic moment orientations of the two sublattices. $I_{DS}$ flows along b-axis. (c)-(d) $I_{DS}$ as a function of the applied field in the subthreshold region ($V_{GS}$ = -5 V) (c), and in the electrical ON region ($V_{GS}$ = 20 V) (d), for the drain current flowing along the b- (top) and a- (bottom) axes. (e) $I_{DS}$ vs. $V_{GS}$ for devices with current flowing along b- (closed squares) and a- (open circles) axes, with (red) and without (gray) magnetic field. The two curves of $I_{DS} \parallel a$ device mostly overlap due to the small magnetoresistance. The aspect ratio $W/L$ for $I_{DS} \parallel a$ device is 10 and that for $I_{DS} \parallel b$ device is 20. (f) Mobilities, $\mu_{n,a}$ (open circles) and $\mu_{n,b}$

(closed squares), with (red) and without (gray) magnetic field extracted from the linear region of (e). All the data in this figure is taken under 70 K and $V_{DS}$ = 10 V.

Chung-Tao Chou[1,2], Eugene Park[3], Josep Ingla-Aynes[4], Julian Klein[3], Kseniia Mosina[5], Jagadeesh S. Moodera[1,4], Zdenek Sofer[5], Frances M. Ross[3], Luqiao Liu[2*]

[1]Department of Physics, Massachusetts Institute of Technology, Cambridge, MA 02139, USA

[2]Department of Electrical Engineering and Computer Science, Massachusetts Institute of Technology, Cambridge, MA 02139, USA

[3]Department of Materials Science and Engineering, Massachusetts Institute of Technology, Cambridge, MA 02139, USA

[4]Francis Bitter Magnet Laboratory and Plasma Science and Fusion Center, Massachusetts Institute of Technology, Cambridge, Massachusetts 02139, USA

[5]Department of Inorganic Chemistry University of Chemistry and Technology Prague, Technická 5, Prague 166 28, Czech Republic

*Corresponding author. Email: luqiao@mit.edu

**Table of contents:**

**Experimental Methods**

**Synthesis of CrSBr**

CrSBr bulk single crystal was synthesized by a chemical vapor transport method. Chromium (99.99%, -60 mesh, Chemsavers, USA), sulfur (99.9999%, 1 - 6 mm Wuhan Xinrong New Materials Co., China), and bromine (99.9999%, Merck, Czech Republic) elements were placed with a stoichiometry of 1:1:1 in a quartz ampoule (60x250mm). The amounts corresponded to 30g of CrSBr using 2 at.% excess of Br and S. The ampoule, cooled with liquid nitrogen, was melt sealed under high vacuum using an oil diffusion pump with liquid nitrogen trap. The ampoule was then heated in a crucible furnace at 400 °C for 50 hours, at 500°C for 50 hours and at 600°C for 50 hours. The top part of the ampoule was kept under 100 °C to avoid excessive pressure formation in the ampoule. The reacted ampoule was placed in a two zone furnace, where first the source zone was heated to 700 °C and the growth zone to 900 °C. After two days the thermal gradient was reversed and the source zone was kept at 900 °C and growth zone at 800 °C for 10 days. Finally, the ampoule was cooled to room temperature and opened inside an argon-filled glovebox.

**Device fabrication**

Heavily p-doped silicon substrates (0.001-0.005 ohm-cm) with 90 nm dry chlorinated and forming gas annealed thermal oxide were used for CrSBr MOSFET devices. The thermal oxide was annealed with forming gas to passivate defects at the top surface [S1]. Micrometer-sized Ta(5 nm)/Pt(5 nm) source and drain were deposited on the gate oxide by dc magnetron sputtering followed by patterning with photolithography and a lift-off process. To remove sidewalls or ribbons that formed during the lift-off process, the substrates were sonicated in acetone for 3 minutes. An oxygen plasma asher was used to remove organic residues after the sonication. CrSBr grown by chemical vapor transport was mechanically exfoliated in ambient air, and bulk-like flakes with smooth surface and thickness in the range ranging from tens of nm to a few hundred nm were transferred onto the pre-patterned source and drain electrodes. The flake is aligned with the source and drain so that the channel current flows in b-axis direction unless otherwise specified. Before this transfer, both CrSBr and oxide surfaces were inspected under an optical microscope to ensure the flake had no cracks and the processing had not left metal ribbon or sidewalls on the oxide. After the transfer, each device was heated to 80 °C for 5 minutes to enhance adhesion. Images of finished devices are shown in Fig. S1(a)-(c). Several aspects of the fabrication process were optimized to reduce interfacial contamination. Use of lift-off rather than etching avoids oxide damage due to over-etching. A metal-ion free developer was used in photolithography. Flakes were transferred using a dry process rather than a solvent-based process to prevent trapping of contamination between CrSBr and oxide. We also found that our MagFET devices for which the CrSBr transfer process took place in air consistently showed superior performance compared to those for which the transfer process took place in an $N_2$ glove box. Transfer in air results in mild surface oxidation of CrSBr [S2] and hydrolysis of Br that may lower $D_{it}$.

**X-ray Diffraction**

X-ray powder diffraction data measurements were conducted under the ambient conditions on Bruker D8 Discoverer (Bruker, Germany) powder diffractometer with parafocusing Bragg–Brentano geometry and CuKα radiation ($\lambda$ = 0.15 418 nm, U = 40 kV, I = 40 mA). Data were collected over the angular range 5–90° (2θ) with a step size of 0.019° (2θ). The acquired data were analyzed by using HighScore Plus 3.0 software.

**Scanning Transmission Electron Microscopy Characterization**

Devices were imaged in a cross-sectional geometry, with the cross section cut through the CrSBr crystal at locations over the electrodes and oxide layer. The sections were prepared using a FEI Helios Nanolab 600 Dual Beam System. After inserting a device in the FEI Helios system, protective layers of C then Pt were deposited over the CrSBr region using the electron beam. A cross-section was milled with $Ga^+$ ions, extracted with an Omniprobe nanomanipulator, and attached to a TEM copper half grid (manufacturer Ted Pella) using carbon deposited with the ion beam. Final thinning of the sample was then performed using $Ga^+$ ions to obtain a thickness of approximately 70 nm. Samples prepared in this way were imaged in an aberration-corrected Thermo Fisher Themis Z STEM operated at 200 kV. The probe convergence angle was 18.9 mrad, beam current was 30-40 pA and frame size was 1024 x 1024 pixels with a 500ns/pixel dwell time. Ten to fifteen images were acquired and were drift corrected using the drift correction frame integration (DCFI) function of the Velox software to increase the signal-to-noise ratio.

**Electrical and magnetoresistance measurements**

The temperature-dependent electrical transport measurements were carried out in a Janis cryostat with Lakeshore 336 temperature controller, while all the other transport measurements were carried out in a Quantum Design Physical Properties Measurement System (PPMS). External multimeters (Keithley 2400) are connected to the device for applying gate and drain voltages and measuring source-drain current. Low frequency ac measurements with a lock-in amplifier (Stanford Research 810) were carried out separately and consistent results were observed. A relatively low voltage $V_{DS}$ (3V) was used in the measurement in Fig. S2(a) to avoid any potential local heating. For the other $I_{DS}$-$V_{GS}$ measurements, large $V_{DS}$ (10 V) was used to minimize the contribution of the Schottky barrier or junction resistance at the source and drain contact.

**Vibrating Sample Magnetometer of CrSBr**

The magnetic moment of CrSBr is measured in Quantum Design PPMS with a linear transport motor and a vibrating sample magnetometer coil set. A millimeter sized bulk CrSBr is mounted with b-axis aligned with magnetic field direction for the measurement.

**Finite element simulation of current distribution**

Finite element simulation is carried out by a homemade MATLAB program which accounts for the anisotropic conductance in b-axis and c-axis directions. The simulation assumes 2-dimensional current distribution (in the plane formed by b-axis and c-axis) and divides the device in a 32×256 mesh. See Supplementary Text Section 1 for more information.

**Undoped MOSFET modeling**

The undoped MOSFET device [S3] is modeled by homemade MATLAB program, which calculates the carrier density, electric field, and conduction band minimum as a function of depth inside CrSBr according to one-dimensional self-consistent Boltzmann-Poisson equation. An interface trap density $D_{it}$ is included in the model to account for the reduced subthreshold swing [S4]. See Supplementary Text Section 2 and Fig. S5 for more information.

**Mobility extraction**

The electronic mobility is extracted from $I_{DS}$-$V_{GS}$ curves by performing a linear regression in the electrical ON regime (20V<$V_{GS}$<25V) and obtaining the slope of $I_{DS}$-$V_{GS}$ curves. The mobility is given by [36]:

$$\mu_n = \frac{L}{W} \times \frac{V_{DS}}{C_{ox}} \times \frac{dI_{DS}}{dV_{GS}}.$$

The mobility at lower temperature (T<30 K) cannot be extracted this way since the device does not reach linear region when $V_{GS}$ is up to 25 V due to increased $V_{th}$.

**Interface trap density estimation**

The interface trap density $D_{it}$ is extracted from the subthreshold slope of the $I_{DS}$-$V_{GS}$ curves by the formula:

$$\frac{d}{dV_{GS}} \ln(I_{DS}) = \frac{q_e}{k_B T} \times \frac{C_{ox}}{C_{ox} + q_e D_{it}}.$$

The oxide capacitance $C_{ox}$ is estimated by the SiO$_2$ thickness assuming the air gap between CrSBr and SiO$_2$ is negligible.

**Spin Lattice Energy Calculation**

The spin lattice energies of an 8-layer CrSBr in different magnetic states [Fig. 4(c)] are calculated by:

$$E = \sum_i \boldsymbol{H} \cdot \boldsymbol{S_i} + \sum_{<i,j>} -J\,(\boldsymbol{S_i} \cdot \boldsymbol{S_j})$$

where $-J$ is the interlayer antiferromagnetic exchange, $\boldsymbol{H}$ is the applied magnetic field along the easy axis, $\boldsymbol{S_i}$ is the magnetic orientation for the $i^{th}$ layer ($i$=1~8), and $<i,j>$ stands for adjacent pairs of layers. The intermediate transition behavior for CrSBr with more layers is qualitatively the same as the 8-layer example.

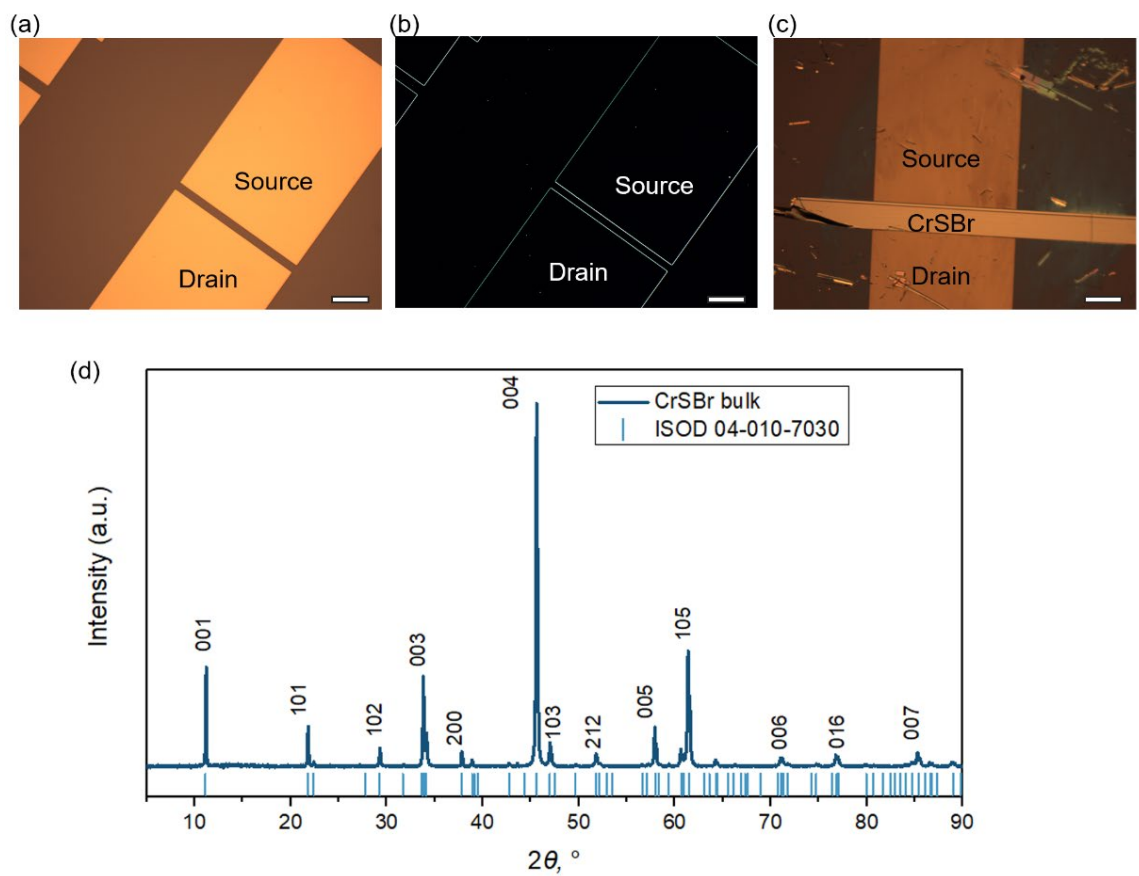

Fig. S1. (a)-(b) Optical image of pre-patterned source/drain before transferring the CrSBr, under bright (a) and dark (b) field. (c) Overview of the magnetic MOSFET. The relatively dirty surface in (c) is due to residual adhesive left behind during the transferring step and is not expected to influence the device performance. Scale bars are 100 μm. (d) X-ray diffraction of CrSBr. The peaks, most of which are in the <k0l> and <00l> directions, are consistent with the orthorhombic bulk CrSBr structure given in PDF 04-010-7030 with Pmmm space group [S5]. The sample used for diffraction is CrSBr powder.

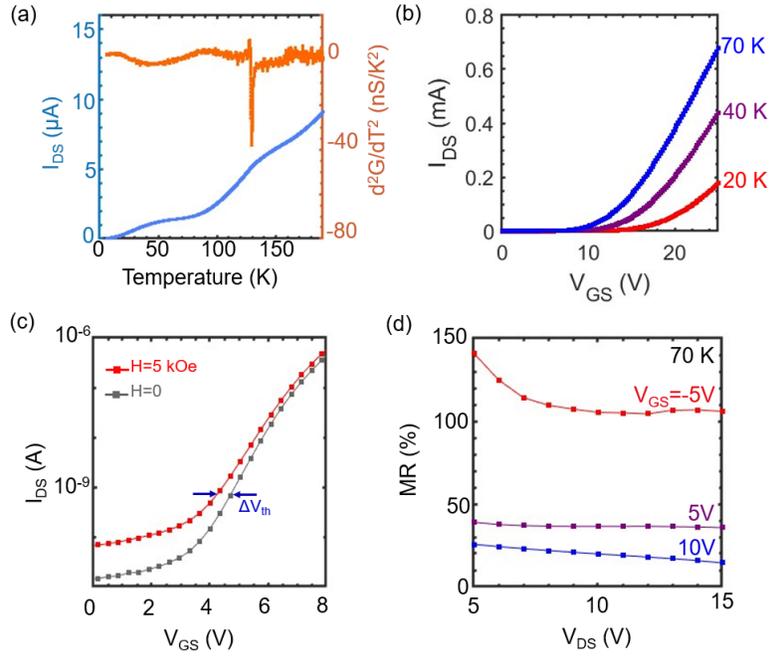

Fig. S2. (a) Temperature dependence of $I_{DS}$ and the corresponding second derivative of the conductance. $V_{DS}$ = 3 V and $V_{GS}$ = 15 V are used in this measurement. (b) Same $I_{DS}$ vs. $V_{GS}$ data as in Fig. 2(a) plotted in linear scale. The threshold voltage increases with decreasing temperature. (c) $I_{DS}$ vs. $V_{GS}$ in the log scale measured at 40 K. (d) $V_{DS}$ dependence of channel MR ratio at 70 K. The MR ratio is mostly constant under large bias voltage ($V_{DS}$>10 V), which is consistent with the threshold voltage shift picture.

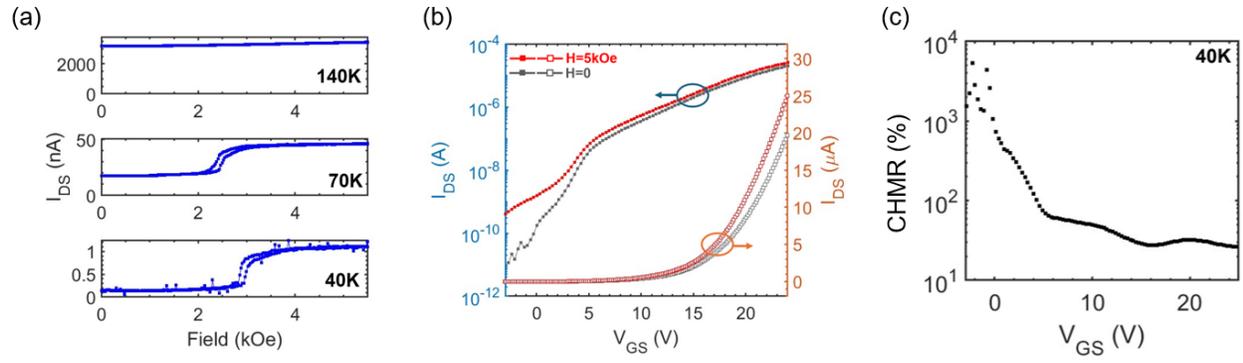

Fig. S3. (a) Magnetoresistance at different temperature. (b) $I_{DS}$-$V_{GS}$ curves in AP and P states. (c) Channel MR versus $V_{GS}$ curve at 40 K. The figures are qualitatively consistent with Fig. 2 in the main text and shows the large MR are repeatable across multiple devices.

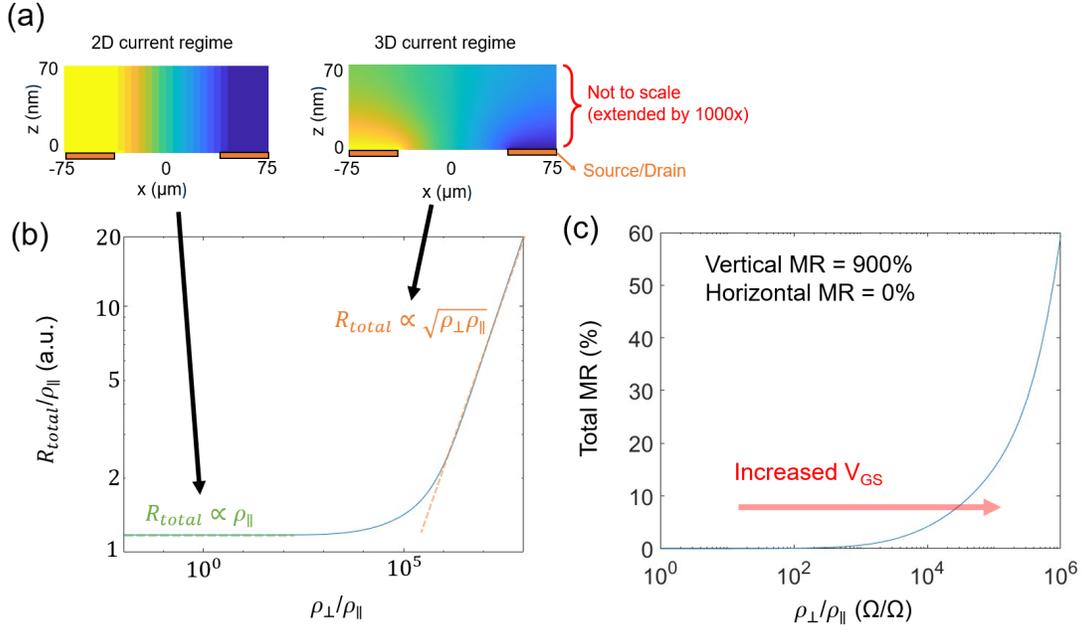

Fig. S4. (a) Schematics of device dimensions and voltage distribution under 2D and 3D conduction regime. (b) Total resistance as a function of ratio of in-plane $\rho_\parallel$ and vertical resistivity $\rho_\perp$. The total resistance follows different asymptotes under low and high $\rho_\perp / \rho_\parallel$. Under low $\rho_\perp / \rho_\parallel$, $R_{total}$ is dominated by the parallel resistance, $R_{total} \propto \rho_\parallel$. For very high $\rho_\perp / \rho_\parallel$, the current distribution along $z$ axis becomes non-uniform and $R_{total}$ depends on both $\rho_\perp$ and $\rho_\parallel$ through $R_{total} \propto \sqrt{\rho_\perp \rho_\parallel}$. Orange lines indicate contact metal region. The dimension along the z-axis is scaled by 1000x to clearly show the device geometry. (c) Magnetoresistance ratio versus conductivity anisotropy, when only junction magnetoresistance is considered. In the simulation, the magnetoresistance (MR) along the vertical direction is assumed to be 900% and in-plane magnetoresistance is set to 0% to show that the vertical magnetoresistance alone cannot explain the total MR measured in the experiment.

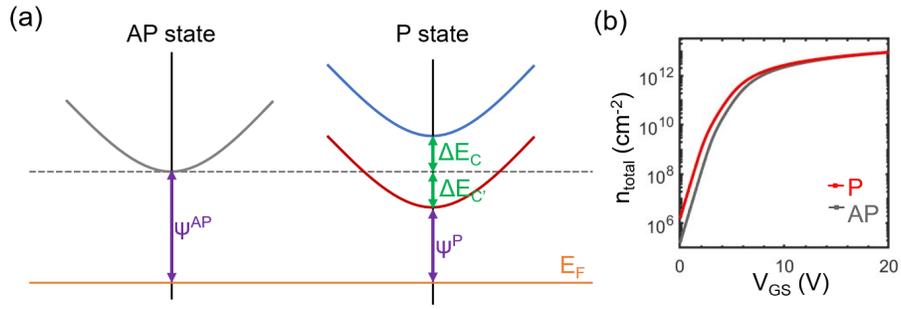

Fig. S5. (a) Schematics of a simplified band structure. The conduction band in the AP state is two-fold spin degenerate while the two spin sub-bands in the P state are split by $\Delta E_C + \Delta E_{C'}$. (b) Calculated $n_{total}$-$V_{GS}$ curves of a magnetic MOSFET in P and AP states. The simulation is done by setting $\Delta E_C = \Delta E_{C'} = 9$ meV and temperature at 35 K.

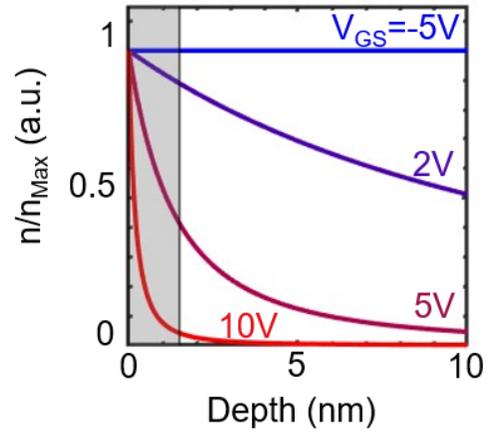

Fig. S6. Distribution of carrier density in CrSBr as a function of depth. The depth is defined as the distance from the CrSBr/SiO$_2$ interface, calculated using self-consistent Boltzmann-Poisson equation described in Supplementary Text.

**Supplementary Text**

**Section 1. Finite element simulation of magnetoresistance due to vertical and in-plane magnetic response.**

We carry out finite element simulation to understand the origin of the total magnetoresistance demonstrated in the transistor device by considering the contributions from resistances along both the in-plane ($x$) and the out-of-plane ($z$) directions.

The device geometry used for the simulation is illustrated in Fig. S4(a), which is very close to the dimension of the device shown in Fig. S3. To account for the differences of resistivities along the $x$ and $z$ direction, we consider a wide range of $\rho_\perp/\rho_\parallel$ ratios. To show that the magnetoresistance along the vertical direction (the tunneling or hopping magnetoresistance) cannot account for the observed phenomena, in the simulation we only assume $\rho_\perp(H)$ exhibits a magnetoresistance while $\rho_\parallel$ remains to be irresponsive to magnetic field. We calculate the voltage and current distributions under each $\rho_\perp/\rho_\parallel$ value. The resultant total resistance of the device $R_{\text{total}} = V_{\text{DS}}/I_{\text{DS}}$ is shown in Fig. S4(b). When the in-plane resistance dominates ($\rho_\perp/\rho_\parallel$ being low), the current distributes more uniformly across the $z$ (film normal) direction, and the voltage drop along the vertical direction is negligible. The total resistance in this limit is largely independent on $\rho_\perp$. Correspondingly, the magnetoresistance along the $z$ direction does not make noticeable contributions to the total magnetoresistance. In Fig. S4(c), we see that the total magnetoresistance remains nearly zero in the low $\rho_\perp/\rho_\parallel$ range. On the other hand, when $\rho_\perp/\rho_\parallel$ becomes much larger, the current distribution is non-uniform along the $z$-direction and the change in $\rho_\perp(H)$ starts to make contributions. In this case, the total resistance scales as $R_{\text{total}} \propto \sqrt{\rho_\perp \rho_\parallel}$, as illustrated in the high $\rho_\perp/\rho_\parallel$ end of Fig. S4(b). Correspondingly, the total magnetoresistance shows a weak dependence on the vertical magnetoresistance. In Fig. S4(c), we assume the magnetoresistance along the vertical direction $MR_\perp$ to be 900%, a high value for the measurement temperature of 30 K based on previous reports [25], and zero magnetic response along the in-plane direction. We see that even in this limit the total magnetoresistance $MR_{total}$ has a much smaller value compared with the original magnetoresistance in the vertical direction ($MR_{total} \leq \sqrt{1 + MR_\perp} - 1 \approx 210\%$), which is not enough to account for the magnetoresistance observed in our experiment (1500%).

In addition, in Fig. S4(c) we see that as the gate voltage increases and the in-plane resistance reduces in the subthreshold regime ($\rho_\perp/\rho_\parallel$ gets larger), the total magnetoresistance would increase when considering only the magnetic response along the vertical direction, which is inconsistent with the $V_{GS}$ dependence measurement in Fig. 3(a) of main text.

**Section 2. Undoped MOSFET modeling**

The CrSBr transistor can be modeled with undoped junctionless MOSFET theory, as the background dopant density is low and a majority of the dopants freeze out at <60K, as evident by the high on-off ratio of up to $10^7$. To account for the magnetic modulation of the electronic band structure, we use a minimalistic model where the two spin sub-bands are degenerate in the AP state and split by $\Delta E_C + \Delta E_{C\prime}$ in the P state [Fig. S5(a)]. The carrier density in the conduction band in AP and P states are given by:

$$n^{\text{AP}}(x) = 2n_i e^{\frac{-q_e \psi^{\text{AP}}(x)}{k_B T}} \quad (1),$$

$$n^{\text{P}}(x) = n_i e^{\frac{-q_e(\psi^{\text{P}}(x) - \Delta E_C)}{k_B T}} + n_i e^{\frac{-q_e(\psi^{\text{P}}(x) + \Delta E_{C\prime})}{k_B T}} \quad (2),$$

where $n^{\text{AP,P}}(x)$ is the conduction band electron density in the AP and P case, respectively, $q_e$ is the electron charge, $\psi^{\text{AP,P}}(x)$ is the energy difference from the Fermi surface to the conduction band minimum [Fig. S5(a)], $\Delta E_C$ is the conduction band edge modulation given by $\Delta E_C = E_C^{\text{AP}} - E_C^{\text{P}}$, and $n_i$ is the intrinsic carrier density of CrSBr given by $n_i = \sqrt{N_C N_V} e^{\frac{-E_g^{\text{AP}}}{2k_B T}}$, where $N_C$ and $N_V$ are the effective density of states for conduction and valence band, respectively.

Since the channel length of the device is large (20 µm), the lateral electric field near source and drain is neglected, leaving the band bending inside CrSBr given by the 1D Poisson equation:

$$\frac{d^2\psi^{AP,P}(x)}{dx^2} = \frac{-q_e}{\epsilon_{CrSBr}} n^{AP,P}(x), \qquad (3)$$

where $\epsilon_{CrSBr}$ is the permittivity inside CrSBr. Since the channel length (20 µm) is orders of magnitude larger than the CrSBr flake thickness (~100 nm), a boundary condition of zero electric field at the far end of the flake is assumed for both on and off regimes:

$$\frac{d\psi^{AP,P}(L)}{dx} = 0, \qquad (4)$$

where $L$ is the thickness of CrSBr. Here, $x=0$ is defined as the semiconductor/oxide interface and $x=L$ is the opposing side of the CrSBr. Analytic solution of Eq. (1), (2), and (3) with the boundary condition of Eq (4) is given by:

$$\frac{q_e}{2k_BT}\left(\psi^{AP}(x) - \psi^{AP}(L)\right) = -\ln\left(\cos\left(\sqrt{\frac{2q_e^2 n_i}{2\epsilon_{CrSBr} k_B T}} e^{\frac{q_e\psi^{AP}(L)}{2k_BT}}(x-L)\right)\right), \qquad (5)$$

$$\frac{q_e}{2k_BT}\left(\psi^{P}(x) - \psi^{P}(L)\right) = -\ln\left(\cos\left(\sqrt{\frac{\left(e^{\frac{\Delta E_C}{k_BT}} + e^{\frac{-\Delta E_{C'}}{k_BT}}\right)q_e^2 n_i}{2\epsilon_{CrSBr} k_B T}} e^{\frac{q_e\psi^{P}(L)}{2k_BT}}(x-L)\right)\right), \qquad (6)$$

The electric field at $x=0$ inside CrSBr is given by the integration of $n(x)$:

$$E_{CrSBr}(x=0) = \frac{q_e}{\epsilon_{CrSBr}} \int_0^L n(x)\, dx, \qquad (7)$$

while the electric field inside the SiO2 is given by:

$$E_{SiO2} = \frac{\epsilon_{CrSBr}}{\epsilon_{SiO2}} E_{CrSBr}(x=0) + \frac{q_e}{\epsilon_{SiO2}} D_{it}(\psi(x=0) - \psi_0), \qquad (8)$$

where $\epsilon_{SiO2}$ is the permittivity of SiO2, $D_{it}$ is the interface trap density of states at the semiconductor/oxide interface, and $\psi_0$ is the chemical potential at which the interface trap is charge neutral. The introduction of the $D_{it}$ term is commonly used in literature to capture the device behavior where the subthreshold swing is reduced by thick gate oxide [36]. Finally, the gate voltage $V_{GS}$ corresponding to $\psi(x)$ is given by:

$$V_{GS} = \psi(0) + E_{SiO2} t_{SiO2} \quad (9),$$

where $t_{SiO2}$ is the thickness of oxide (90 nm). Then, for each given $V_{GS}$, the total 2-dimentional carrier density $n_{total}^{AP}$ and $n_{total}^{P}$ are calculated:

$$n_{Total}^{AP,P} = \int_0^L n^{AP,P}\, dx \qquad (10),$$

The ratio of $n_{Total}$ and ratio of mobility $\mu_n$ for AP and P state gives the MR ratio:

$$\frac{I_{DS}^P}{I_{DS}^{AP}} = \frac{\mu_n^P}{\mu_n^{AP}} \times \frac{n_{total}^P}{n_{total}^{AP}} (11),$$

The calculated $n_{total}^{AP}$ and $n_{total}^{P}$ are shown in Figure S5(b). In the subthreshold regime, where the band bending is negligible due to low carrier density, we have:

$$\psi(x) \sim \psi_{const}, (12)$$

$$n^{AP,P}(x) \sim n_{const}^{AP,P}, (13),$$

$$n_{total}^{AP,P} \sim n_{const}^{AP,P} L, (14)$$

Combining Eq. (1), (2), (11), (14), we arrive at the subthreshold MR given by:

$$\frac{I_{DS}^P}{I_{DS}^{AP}} = \frac{\mu_n^P}{2\mu_n^{AP}}(e^{\frac{\Delta E_C}{k_B T}} + e^{\frac{-\Delta E_{C'}}{k_B T}}) \approx \frac{\mu_n^P}{2\mu_n^{AP}} e^{\frac{\Delta E_C}{k_B T}}. \qquad (15)$$

Here, Eq. (15) is the same as Eq. (1) in the main text.

**Supplementary Reference**

[1] Onishi, K. *et al.* Improvement of surface carrier mobility of HfO$_2$ MOSFETs by high-temperature forming gas annealing. *IEEE Transactions on Electron Devices* **50**, 384–390 (2003).

[2] Torres, K. et al. Probing defects and spin-phonon coupling in CrSBr via resonant Raman scattering. *Advanced Functional Materials* **33**, 2211366 (2023).

[3] Taur, Y. An analytical solution to a double-gate MOSFET with undoped body. *IEEE Electron Device Letters* **21**, 245–247 (2000).

[4] Chen, J. *et al.* Investigation on the interface trap characteristics in a p-channel GaN MOSFET through temperature-dependent subthreshold slope analysis. *Journal of Physics D: Applied Physics* **55**, 095112 (2021).

[5] Data retrieved from Crystallography Open Database 04-010-7030, http://www.crystallography.net/cod/